\documentclass[aps,prd,twocolumn,groupedaddress,showpacs]{revtex4}
\usepackage{epsf,epsfig,graphics,graphicx}
\begin{document}

\title{Bound on the time variation of the fine structure constant\\
driven by quintessence}

\author{Da-Shin Lee$^1$, Wolung Lee$^2$, and Kin-Wang Ng$^2$}
\affiliation{$^1$Department of Physics, National Dong Hwa University,
Hua-Lien, Taiwan 974, R.O.C.\\
$^2$Institute of Physics, Academia Sinica, Nankang, Taipei,
Taiwan 115, R.O.C.}

\begin{abstract}
The bound on the time variation of the fine structure constant
($\alpha$) driven by the dynamics of quintessence scalar field
which is coupled to electromagnetism is discussed using
phenomenological quintessential models constrained by SNIa and CMB
observations. We find that those models allowing early
quintessence give the largest variation $\Delta\alpha$ at the
decoupling epoch. Furthermore, the fifth force experiments imply
that $\Delta\alpha/\alpha$ is less than about $0.1\%$.
\end{abstract}

\maketitle

The idea of time-varying fundamental physical constants such as
gauge coupling constants was proposed long time ago~\cite{dirac}.
Recently it has been claimed that the results of a search for
time variability of the fine structure constant ($\alpha$), using
absorption systems in the spectra of distant quasars, yield
5-$\sigma$ evidence for a smaller $\alpha$ in the past:
$\Delta\alpha/\alpha = -0.543 \pm 0.116 \times 10^{-5}$ over the
redshift range $0.2 < z < 3.7$~\cite{webb}. In addition, there
exist terrestrial and cosmological constraints.
An analysis of the isotropic abundances in the
Oklo natural uranium fission reactor, active about $1.8\times
10^9$ years ago (corresponding to $z\simeq 0.14$), suggests
$-0.9\times 10^{-7}<\Delta\alpha/\alpha<1.2\times
10^{-7}$~\cite{oklo}. A bound $\Delta\alpha/\alpha<3\times
10^{-7}$ has been obtained from the analysis of the Re/Os ratio in
meteorites ($z\simeq 0.45$)~\cite{olive}. While the data of cosmic
microwave background (CMB) anisotropies are consistent with
$\alpha$ being smaller by a few percent at the decoupling epoch
($z\sim 1100$) in a flat cold dark matter model with a cosmological
constant ($\Lambda$CDM)~\cite{bat}, the recent observations made
by the Wilkinson Microwave Anisotropy Probe (WMAP) have provided a
bound $-0.06<\Delta\alpha/\alpha<0.02$ at 95\% CL~\cite{mar03}. At
much higher redshifts, big bang nucleosynthesis considerations
place bounds $|\Delta\alpha/\alpha|<10^{-2}$ at $z\sim
10^9$~\cite{bbn}.

The recent astrophysical and cosmological observations such as
type Ia supernovae (SNe) and CMB anisotropies concordantly prevail
the $\Lambda$CDM model~\cite{ctrig}. The current data, however,
are consistent with a somewhat broader diversity of such
``dark energy'' as long as its equation of state (EOS) approaches that of the
cosmological constant at a recent epoch. A dynamically evolving scalar
field $\phi$ called ``quintessence'' (Q) is probably the most popular
scenario so far to accommodate the dark energy component. Many Q
models have been proposed with various physically motivated effective
potentials $V(\phi)$ for the scalar field~\cite{qmods}.
Several attempts have been made to test different Q
models~\cite{many}. Nevertheless, it proves to be premature at
this stage to perform a meaningful data fitting to a particular Q
model, or to differentiate between the variations. The
reconstruction of $V(\phi)$ would likely require next-generation
observations. As such, model independent approaches which simply involve
parametrizing the dark energy EOS have been proposed to
comply with the SNe and CMB observational data~\cite{lin,cor,lee03}.

It is of particular interest to study the
observational effects of direct interaction of $\phi$
to ordinary matter if there is any. For example,
imposing an approximate global symmetry would allow a coupling of
$\phi$ to electromagnetism, which would lead to rotation of polarized light
from distant radio sources, temporal evolution of
$\alpha$~\cite{carr}, and generation of primordial magnetic
fields~\cite{lee02}. Recently, there have been many studies on the
time-varying $\alpha$ in the context of quintessential cosmology
by invoking non-renormalizable $\phi$-photon
couplings~\cite{qalpha,anchor,park}. However, most of the
studies are based on certain Q models~\cite{anchor}. In
this paper, we will discuss the observational constraints
to the evolution of $\phi$ through its EOS $p_\phi = w_\phi
\rho_\phi$ in the parametrized Q
models~\cite{lin,cor,lee03} that in turn give rise
to a model independent bound on the Q-driven time variation of
$\alpha$ at the decoupling epoch.

The $\phi$-photon coupling that we consider here is
\begin{equation}
L_{\phi \gamma} = -{\kappa\over 4} \frac{\phi}{M_p}
                  g^{\alpha\mu}g^{\beta\nu} F_{\alpha\beta} F_{\mu\nu},
\end{equation}
where $F_{\mu\nu}=\partial_\mu A_\nu - \partial_\nu A_\mu$, $M_p$ is the
reduced Planck mass $(8\pi G)^{-1/2}$, and $\kappa$ is a coupling
constant. The fifth-force experiments limit~\cite{pos}
\begin{equation}
\kappa<10^{-3}.
\label{ff}
\end{equation}
Therefore, by defining $\theta\equiv\phi/M_p$, the change of $\alpha$ from
the present time to a time $t$ is simply given by
\begin{equation}
\frac{\Delta\alpha}{\kappa\alpha}= \theta_0-\theta(t).
\end{equation}
Consider a flat universe in which the total density parameter of the Universe
today is represented by $\Omega_0 = \Omega_m^0+\Omega_r^0+\Omega_\phi^0 = 1$
with $\Omega_m^0h^2 = 0.135$ and $\Omega_r^0h^2 = 4.152\times 10^{-5}$,
where the present Hubble constant is parametrized as
$H_0=100 h {\rm km s^{-1} Mpc^{-1}}$.
Assuming a spatially homogeneous $\phi$ field, the evolution of the cosmic
background is governed by
\begin{eqnarray}
{d\varrho_\phi}/{d\eta}&=& -3a{\mathcal H} \left(1+w_\phi\right)
                               \varrho_\phi,
\label{bkg1} \\
{d{\mathcal H}}/{d\eta}&=& -{3\over2}a{\mathcal H}^2-
{1\over2}a(w_r\varrho_r+w_\phi\varrho_\phi),
\label{bkg2}
\end{eqnarray}
where $a$ is the scale factor and the conformal time is defined as
$\eta=H_0\int dt a^{-1}(t)$. Here we have used
$(d\phi/dt)^2=(1+w_\phi)\rho_\phi$ and
$V(\phi)=(1-w_\phi)\rho_\phi/2$, and rescaled the energy density
of the $i$th component as $\varrho_i \equiv \rho_i/(M_pH_0)^2$.
Accordingly, the dimensionless Hubble parameter is given by
\begin{equation}
    {\mathcal H}^2\equiv (H/H_0)^2 =
    \Omega_m^0~a^{-3}+\Omega_r^0~a^{-4}+\Omega_\phi~{\mathcal H}^2.
\end{equation}
Therefore, given the prescribed EOS $w_\phi(\eta)$
which is a function of conformal time, the problem is reduced to
solving a set of first-order coupled ordinary differential
equations (\ref{bkg1}) and (\ref{bkg2}) with the initial
conditions set at the present time and the time variable running
backward. Let us introduce a useful quantity
that is the $\Omega_\phi$-weighted average~\cite{huey99}
\begin{equation}
\langle w_\phi \rangle \equiv \int_{\eta_{\rm dec}}^{\eta_0}
\Omega_\phi(\eta) w_\phi(\eta) d\eta
\times \left( \int_{\eta_{\rm dec}}^{\eta_0}
\Omega_\phi(\eta) d\eta \right)^{-1},
\end{equation}
where $\eta_0$ and $\eta_{\rm dec}$ are the conformal time today
and at decoupling respectively. It is known that $w_\phi\neq 0$
after decoupling would result in a time-varying Newtonian
potential which produces large-scale CMB anisotropy through the
integrated Sachs-Wolfe (ISW) effect. It was shown~\cite{huey99,bean}
that as far as the CMB anisotropy power spectrum is concerned, the
time-averaged $\langle w_\phi \rangle$ can be well approximated by
an effective constant value for $w_\phi$ as long as the Q
component is negligible at decoupling.

Physically, $-1\le w_\phi\le 1$, where the former
equality holds for a pure vacuum state. Lately some progress has
been made in constraining the behavior of $\phi$ from
observational data. A combined large scale structure, SNe, and CMB
analysis has set an upper limit on Q models with a constant
$w_\phi<-0.7$~\cite{bond,bean}, and the recent WMAP CMB data gives
a stronger limit $w_\phi<-0.78$~\cite{wmap}. Furthermore, the SNe
data and measurements of the position of the acoustic peaks in the
CMB anisotropy spectrum have been used to put a constraint on the
present $w_\phi^0 \le -0.96$~\cite{cope}. The apparent brightness
of the farthest SN observed to date, SN 1997ff at redshift $z\sim
1.7$, is consistent with that expected in the decelerating phase
of the flat $\Lambda$CDM model with $\Omega_\Lambda \sim
0.7$~\cite{riess}, implying $w_\phi= -1$ for $z<1.7$.

Let us first work on the simplest case with a constant $w_\phi=-0.8$
to obtain an order-of-magnitude estimate on
$\Delta\alpha/\kappa\alpha$. By solving
\begin{equation}
d\theta/d\eta=a{\sqrt{(1+w_\phi){\varrho}_\phi}},
\label{dtheta}
\end{equation}
we find that $\Delta\alpha/\kappa\alpha$ decreases quickly with redshift
to a constant value of $-0.82$ at $z\sim 20$.
Next we study the varying $\alpha$ in two phenomenological Q models
which are based on the maximum likelihood analysis of the current
SNe and CMB observational data. The first model has a monotonic EOS
$w_\phi(a)=w_0+(w_m-w_0)(1-a)$
(where $w_\phi\rightarrow w_m$ as $a\rightarrow 0$)
that has been shown to be able to describe the
late-time $\phi$ evolution for a wide class of Q models~\cite{lin}.
Using $w_0=-0.99$ and $w_m=-0.31$, which are within $2\sigma$ from
the best-fit parameters to the SNe and CMB data~\cite{cald},
we find that $\Delta\alpha/\kappa\alpha$ decreases to $-1.63$ at decoupling.
The second model has an EOS with four parameters
$W=(w_0,w_m,a_c,\Delta)$~\cite{cor}:
\begin{equation}
w_\phi(a)=w_0+(w_m-w_0)
\frac{\left(1+e^{a_c/\Delta}\right) \left(1-e^{(1-a)/\Delta}\right)}
{\left(1+e^{(a_c-a)/\Delta}\right) \left(1-e^{1/\Delta}\right)},
\end{equation}
where $a_c$ signifies the transition epoch and $\Delta$ represents
the transition rate from $w_0$ to $w_m$. We have worked out two
cases: a slow EOS with $W=(-1,-0.3,0.1,0.8)$ and a rapid EOS with
$W=(-1,0,0.4,0.1)$, both of which are taken from the $1\sigma$
values of the best-fit models to the WMAP and SNe
data~\cite{kunz}. We find that $\Delta\alpha/\kappa\alpha$
decrease to $-1.56$ and $-3.82$ at decoupling for the slow and
rapid EOS respectively. In fact, the slow EOS and the first model
are very similar, in which $\Omega_\phi/\Omega_m$ decreases to
$0.01$ at $z\sim 30$ and becomes negligible in the dark ages. On
the contrary, the rapid EOS which has a larger change in $\alpha$
allows early quintessence with $\Omega_\phi/\Omega_m\sim 0.1$ for
$1100>z>10$. This can be understood by considering a constant
$w_\phi$ for which the problem has a simple solution.
Equation~(\ref{bkg1}) gives ${\varrho}_\phi\propto a^{-3(1+w_\phi)}$
and $\Omega_\phi/\Omega_m \propto a^{-3w_\phi}$. Thus, from
Eq.~(\ref{dtheta}) we have $d\theta/d\eta\propto a^{-(1+3w_\phi)/2}$ 
and therefore the change of $\alpha$ will grow
if $w_\phi\ge -1/3$. Also, the condition for early quintessence is
simply $w_\phi\ge 0$. From the results of the above slow and rapid
EOS's where $w_m=-0.3$ and $w_m=0$ respectively, we see that the
change of $\alpha$ in models with $0>w_\phi>-1/3$ is smaller than
that in models with early quintessence. Therefore, early
quintessence is a crucial condition for a large temporal change of
$\alpha$. We thus see that within viable models the varying
$\alpha$ strongly depends on the Q component in the dark ages.

To further study the Q dependence and to find out the maximum
change of $|\Delta\alpha/\kappa\alpha|$, we turn to the generic
quintessence (GQ) model~\cite{lee03}, in which a piecewise
constant EOS subject to observational constraints is introduced.
The GQ model starts with determining the form for $w_\phi$ at low
redshifts. We choose $w_\phi\simeq -1$ for $z\lesssim 2$ to
satisfy the above-mentioned SNe constraints. Then $w_\phi$ should
be increased to $1$ after $z=2$ to have a maximum change of
$\alpha$ based on Eq.~(\ref{dtheta}). However, to avoid an
unacceptably large ISW effect, $w_\phi$ must drop to $-1$ at a
certain redshift. This results in the square-wave EOS in the GQ model.
\begin{figure}
\leavevmode \hbox{ \epsfxsize=3.2in \epsffile{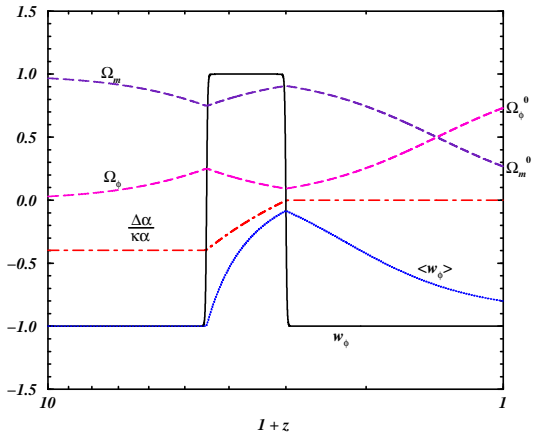}}
\caption{$\Delta\alpha/\kappa\alpha$ in the GQ model with a
square-wave EOS located at low redshift.} \label{fig2}
\end{figure}
A GQ model is shown in Fig.~\ref{fig2}, where
the width of the square wave is determined by $\langle w_\phi\rangle= -0.8$ to
saturate the WMAP upper limit on a constant $w_\phi$~\cite{wmap}. In this case,
$\Delta\alpha/\kappa\alpha$ reaches a minimum value of $-0.4$.
In fact, a square-wave or pulse-shaped EOS is anticipated and quite
general in the class of Q models using pseudo Nambu-Goldstone boson fields
incorporated with quantum corrections to the cosine potential~\cite{corm}.
However, one can still shift the whole square wave to higher redshifts
while fixing $\langle w_\phi \rangle= -0.8$ as to generate the required
ISW effect. We find that the shift increases the Q component
at the decoupling epoch and affects the location
of the CMB acoustic peaks. This can be understood as follows.

The tightly coupled baryon-photon plasma experienced
a serial acoustic oscillation just before the recombination epoch.
The acoustic scale (the angular momentum scale of the acoustic
oscillation) sets the location of the peaks in the
CMB anisotropy power spectrum~\cite{hu95}, and is characterized by
\begin{equation}
    l_A = \pi d_*/h_s =
    \pi (\eta_0-\eta_{\rm dec})/\int_0^{\eta_{\rm dec}} c_s d\eta~,
\label{la}
\end{equation}
where $d_*$ represents the comoving distance to the decoupling epoch
and $h_s$ denotes the sound horizon at decoupling, both of which
are affected in the presence of quintessence.
The sound speed $c_s$ in the pre-recombination plasma is given by~\cite{hu95}
\begin{eqnarray}
    c_s &=& 1/\sqrt{3(1+R)}~~~{\rm with}~~\nonumber \\
    R&\equiv&\frac{3\rho_b}{4\rho_\gamma}\approx 30366\left(
    \frac{T_\gamma^0}{2.725K}\right)^{-4} \frac{\Omega_b^0 h^2}{1+z}~,
\end{eqnarray}
where the baryon-photon momentum density ratio $R$ sets the baryon loading
to the acoustic oscillation of CMB. Hence, $h_s$ at decoupling
can be determined by the differential equation
\begin{equation}
    dh_s/d\eta=c_s,
\end{equation}
coupled with the background evolution equations (\ref{bkg1}) and
(\ref{bkg2}). Using $\Omega_b^0h^2 = 0.0224$, $h = 0.71$,
$T_\gamma^0= 2.725K$, and $z_{\rm dec} = 1089$, we can calculate
various acoustic scales by shifting the square wave
of $w_\phi$ toward the decoupling epoch while keeping
$\left\langle w_\phi\right\rangle$ fixed.%
\begin{figure}
\leavevmode
\hbox{
\epsfxsize=3.2in
\epsffile{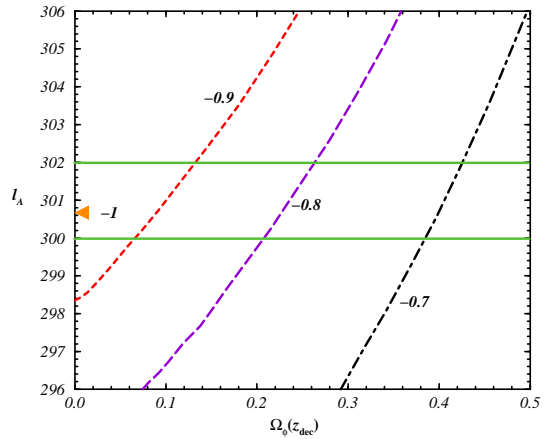}}
\caption{Acoustic scales of the GQ models are plotted as a function of the
quintessence density at decoupling while keeping
$\left\langle w_\phi\right\rangle$ fixed. The model equivalent to the
cosmological constant case has $l_A \simeq 301$ and is denoted by an arrow.
The two horizontal lines signify the 1-$\sigma$ upper and lower bounds
permitted by the WMAP data~\cite{wmap}.}
\label{fig3}
\end{figure}
%
\begin{figure}
\leavevmode
\hbox{
\epsfxsize=3.2in
\epsffile{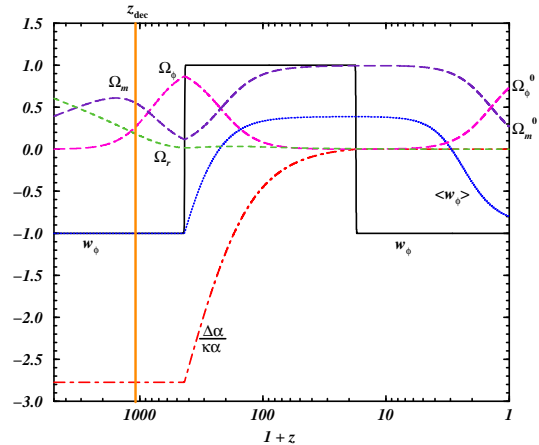}}
\caption{$\Delta\alpha/\kappa\alpha$ in the extreme GQ model with
($\left\langle w_\phi \right\rangle,~\Omega_\phi^{\rm dec}$) = ($-0.8,~0.26$).
The decoupling epoch is marked as the vertical line at $z_{\rm dec} = 1089$.}
\label{fig4}
\end{figure}
Figure~\ref{fig3} plots the results against the Q field energy
density at decoupling $\Omega_\phi^{\rm dec}$ with contours
$\left\langle w_\phi\right\rangle = -0.7$, $-0.8$, $-0.9$, and
$-1$. The last one is simply the $\Lambda$CDM model. All GQ models
lying in the region above the curve of $\left\langle w_\phi
\right\rangle = -0.8$ and within the WMAP bounds are consistent
with the current CMB data. Thus, the constant $w_\phi=-0.8$ model
and that in Fig.~\ref{fig2} are in fact disfavored as far as the
acoustic peak location is concerned. In Fig.~\ref{fig4}, we have
worked out an extreme model with parameters taken at the upper
right corner of the allowed region in Fig.~\ref{fig3}. The model
shows that the Q component dominates over the matter
density at high redshifts. This early quintessence drives
$\Delta\alpha/\kappa\alpha$ to a minimum value of $-2.77$.
Note that a non-zero $\Omega_\phi$ after decoupling would lead to
a suppression of the growth of matter perturbations on scales
smaller than the smoothing scale of the quintessence. Following
Ref.~\cite{lee03}, we have found that the growth function for
matter perturbations in the extreme GQ model is still at an
acceptable level with the COBE data. Furthermore, perturbations in
the dark energy component will mildly affect our results because
dark energy with $\langle w_\phi \rangle= -0.8$ is already close
to a vacuum state~\cite{dave}. At last, the extreme GQ model with
$\Omega_\phi^{\rm dec}=0.26$ is consistent with the upper bound
$\Omega_\phi < 0.39$ during the radiation dominated epoch obtained
by performing a maximum likelihood analysis on the CMB
data~\cite{hansen}.

In conclusion we have investigated the quintessence-induced
evolution of $\alpha$ allowed by the observational constraints
from CMB and SNe using the parametrized quintessence models.
Although the true equation of state, if there is any, may be a
complicated function of time, they should capture the generic
features of quintessence evolution for a model-independent study
of the time-varying $\alpha$.
We have found an extreme GQ model and a rapid equation of state
that allow early quintessence driving a maximum change of
$|\Delta\alpha/\kappa\alpha|\sim 3-4$ at the decoupling epoch.
Hence, the fifth-force limit in Eq.~(\ref{ff}) implies that
$|\Delta\alpha/\alpha|$ is less than about $0.1\%$ at decoupling
as long as the quintessence is coupled to photon. Future CMB data
will be able to constrain $\alpha$ up to $0.1\%$ level or tighter
than this when combined with other 
cosmological measurements~\cite{mar03}.

So far, we have not considered using the phenomenological quintessence models
to fit the observational data of varying $\alpha$ at low redshifts ($z<3.7$).
Although it is {\it ad hoc} and fine-tuned, we find that it is not so difficult
to modify the equation of state at low redshifts to make the change
of $\alpha$ consistent with the measurements of quasar
absorption spectra and the Oklo and Re/Os limits. Indeed, using a
model-independent approach it was shown that these low-redshift measurements
can constrain the dark energy equation of state
today to satisfy $-1<w_\phi<-0.96$ and disfavor late-time changes
in $w_\phi$~\cite{park}. This result is consistent with what we have assumed
for the low-redshift value of $w_\phi$ in our GQ models and the 
constraint $\Delta\alpha/\alpha = -0.06 \pm 0.06 \times 10^{-5}$ recently 
obtained based on a new sample of high-redshift ($0.4\le z\le 2.3$) 
quasar absorption line systems~\cite{sri}.

This work was supported in part by the National Science Council,
Taiwan, ROC under the Grant NSC91-2112-M-001-026 and
NSC91-2112-M-259-008.

\end{document}